\documentclass[preprintnumbers,article,amsmath,amssymb,floatfix,10pt,prd,onecolumn,superscriptaddress,nofootinbib]{revtex4-2}
\usepackage{bm}
\usepackage{amsfonts}
\usepackage{latexsym}
\usepackage[latin1]{inputenc}
\usepackage{graphicx}
\usepackage{amsmath}
\usepackage{palatino}
\usepackage{mathpazo}
\usepackage{textcomp}
\usepackage{textcomp}
\usepackage{physics}
\usepackage{mathrsfs}
\usepackage{float}
\usepackage{booktabs}
\usepackage{dcolumn}
\usepackage{ragged2e}
\usepackage{hyperref}
\usepackage{tensor}
\usepackage[version=3]{mhchem}
\hypersetup{colorlinks,citecolor=blue}
\hypersetup{colorlinks=true,linkcolor=red,filecolor=magenta,    urlcolor=blue}
\usepackage{amsmath}
\usepackage{xcolor}
\usepackage{enumitem}
\usepackage{orcidlink}
\usepackage{epsfig}
\usepackage{subfigure}
\usepackage{commath}
\usepackage{caption}

\begin{document}

\title{Kinematic Probes of Type-II MMG: Pad\'e Cosmographic Analysis of VCDM}
\author{Soumya Kanta Bhoi\orcidlink{0009-0000-4392-317X}}
\email{soumyakanta.1711@gmail.com}
\affiliation{Department of Mathematics, Birla Institute of Technology and Science, Pilani, Hyderabad Campus, Jawahar Nagar, Kapra Mandal, Medchal District, Telangana 500078, India}%
\author{Sai Swagat Mishra\orcidlink{0000-0003-0580-0798}}
\email{saiswagat009@gmail.com}

\affiliation{SR University, Warangal, Telangana, 506371
}%

\author{P.K. Sahoo\orcidlink{0000-0003-2130-8832}}
\email{pksahoo@hyderabad.bits-pilani.ac.in}
\affiliation{Department of Mathematics, Birla Institute of Technology and Science, Pilani, Hyderabad Campus, Jawahar Nagar, Kapra Mandal, Medchal District, Telangana 500078, India}
\begin{abstract}
We study the late-time expansion history of the Universe within the VCDM model, a Type-II MMG realization that preserves the successes of General Relativity while extending beyond constant vacuum energy through a minimal Hamiltonian modification, generating a time-dependent vacuum sector without introducing additional degrees of freedom. We investigate this framework within a cosmographic approach by employing a Pad\'e $P_{(2,1)}$ approximation for the Hubble parameter and luminosity distance, allowing the cosmographic parameters to be expressed directly in terms of the underlying VCDM model parameters and enabling a data-driven reconstruction of the expansion history. The model is constrained within a Bayesian framework using the MCMC technique, implemented via the affine-invariant ensemble sampler, with a joint analysis of cosmic chronometers, DESI BAO, and Type Ia supernova datasets (Union3, Pantheon+, and DESY5). We find that the model parameters are tightly constrained and consistent across different dataset combinations, with the jerk parameter remaining very close to its $\Lambda$CDM value, $j_0 \simeq 1$, indicating no significant deviation at the level of higher-order cosmography. Furthermore, the transition feature previously reported in VCDM is not observed within the Pad\'e $P_{(2,1)}$  cosmographic reconstruction, suggesting that it is not a robust requirement of current observational data but is sensitive to the choice of parametrization. Overall, our results indicate that the VCDM model effectively mimics $\Lambda$CDM at the background level when constrained through a cosmographic approach, underscoring the importance of model-independent reconstructions in assessing alternative cosmological scenarios.
\end{abstract}
\maketitle

\section{Introduction}\label{intro}

Precision cosmology has reached a paradoxical juncture, as the same observations that established $\Lambda$CDM as the standard model of cosmology \cite{SupernovaSearchTeam:1998fmf,SupernovaCosmologyProject:1997czu,Planck2018,Rosenberg:2023ipt} are now exposing its limitations \cite{Bull:2015stt}, from theoretical fine-tuning \cite{Weinberg:1988cp} and the coincidence problem \cite{Zlatev1999} to the statistically significant $\sim 5\sigma$ Hubble tension \cite{Planck2018,Riess2022SH0ES,Verde2019Tension, DiValentino:2021izs,Kavya:2025vsj}. If these discrepancies are not primarily driven by unresolved systematics, they may indicate physics beyond the standard cosmological model, including modified gravity scenarios capable of describing both primordial and late-time cosmic evolution. Modified gravity offers a natural direction, where cosmic acceleration emerges from the gravitational sector itself without invoking a dark energy fluid \cite{Clifton:2011jh,Joyce:2014kja,BeltranJimenez:2019esp, SwagatMishra:2025scq}. In this work, we focus on the VCDM model, which arises as a consistent limit of Type-II Minimally Modified Gravity (MMG) 
\cite{DeFelice:2020eju,Akarsu:2024eoo}. Type-II MMG is distinguished by propagating only the two tensorial degrees of freedom of General Relativity (GR), while still allowing modifications at the cosmological background level \cite{DeFelice:2020eju}. The VCDM realization of Type-II MMG, furthermore, admits a consistent embedding of the $\Lambda_{\rm s}$CDM sign-switching cosmological constant scenario \cite{Akarsu:2024eoo,Akarsu:2024qsi,DeFelice:2020eju}. Recent studies have suggested that this framework can improve the fit to multiple cosmological datasets while potentially easing the tensions $H_0$ and $S_8$ \cite{Akarsu:2024qsi,Arora:2025msq}.  The background Hubble parameter takes a closed-form expression involving a hyperbolic tangent in redshift, smoothly interpolating between matter domination and late-time acceleration \cite{Akarsu:2024eoo}. In its $\Lambda_{\rm s}$VCDM realization, this corresponds to a rapid yet smooth anti-de~Sitter to de~Sitter transition that can potentially alleviate the Hubble tension \cite{Akarsu:2024eoo}. 

To test such deviations with data, we employ cosmography \cite{Weinberg:1972kfs,Visser2004Cosmography, Capozziello:2011hj}, which parametrizes the expansion history solely through successive derivatives of the scale factor, leading to the deceleration $q(z)$ and jerk $j(z)$  without assuming any particular gravity or dark energy model. Among the cosmographic quantities, the jerk parameter is especially informative because the concordance $\Lambda$CDM model predicts $j=1$
\cite{Visser2004Cosmography,Zimdahl:2003wg}, therefore any statistically significant deviation from this value would indicate departures from a constant vacuum energy scenario. Observationally, recent model-independent cosmographic reconstructions based on DESI BAO measurements have indicated statistically significant deviations from the canonical prediction $j_0=1$, with the concordance $\Lambda$CDM scenario exhibiting a moderate $\sim 2\!-\!3\sigma$ tension relative to the cosmographically inferred present-day jerk \cite{Alfano:2025gie,Mishra:2026gbl}. This evidence is further supported by joint analyses including Type Ia supernova compilations such as Pantheon$+$, Union3, and DES-Y5 \cite{Rodrigues:2025tfg}, pointing to the possibility that the late-time expansion history may contain dynamics beyond a strictly constant vacuum energy component.

In the context of VCDM, the cosmographic parameters are not independent phenomenological quantities but must be determined by the underlying model parameters. For the current study, we set $A_3=10^{-3}$ and impose constraints on other parameters employing observational data, as this parameter exhibits a large degeneracy spanning several orders of magnitude and remains unconstrained by current cosmological datasets \cite{DeFelice:2020eju}.  Our methodology is carried out in three steps. First, starting from the exact VCDM Hubble function form \cite{Mishra:2026vdo,DeFelice:2020eju,Akarsu:2024eoo}, we derive closed-form expressions for the present-day cosmographic parameters $q_0$ and $j_0$ \cite{Visser2004Cosmography,Capozziello:2011hj} in terms of $\{\Omega_{m,0}, \beta_H, A_2\}$. Second, these are embedded into a Pad\'e $P_{(2,1)}$ approximant for the luminosity distance $d_L(z)$, yielding a fully model-derived reconstruction. Third, the expansion history $H(z)$ is reconstructed from $d_L(z)$ via differentiation, providing an independent consistency check against observational data. A practical limitation of standard cosmography is that Taylor expansions lose convergence for $z \gtrsim 1$ \cite{Cattoen:2007sk,Vitagliano:2009et}, which is problematic given that modern supernova compilations extend to $z \sim 2.3$ \cite{Brout:2022vxf,Scolnic:2021amr}. To address this issue , we adopt rational approximants Pad\'e $P_{(2,1)}$ \cite{Gruber:2013wua,Wei:2016moy,Hu:2022udt,Mishra:2025vpy,Capozziello:2017ddd}, in which the luminosity distance is expressed as a polynomial ratio rather than a series truncated, which provides the minimal rational form capable of encoding both the deceleration and the jerk parameters while maintaining stability at higher redshifts. In our approach, the coefficients of the Pad\'e $P_{(2,1)}$  approximant are not treated as free parameters but are fixed by matching to the cosmographic expansion. Since the cosmographic parameters themselves are derived analytically from the VCDM background dynamics \cite{Akarsu:2024eoo}, this provides a stable reconstruction at higher redshift that connects the model directly with observable distance measures.

We constrain the model using late-time datasets, including cosmic chronometers \cite{Moresco:2020fbm}, BAO measurements from DESI \cite{DESI:2024mwx}, and Type~Ia supernova samples from Pantheon$+$ \cite{Brout:2022vxf}, DES-Y5 \cite{DES:2024ffp}, and Union3 \cite{Rubin:2023jdq}. The resulting parameter bounds are used to assess the viability of VCDM and its consistency with the $\Lambda$CDM limit. The paper is organized as follows. Section~\ref{sec:vcdm} reviews the VCDM model. Section~\ref{sec:csmgrphy} presents the cosmographic framework and the analytic derivation of the relevant parameters. Section~\ref{sec:pade} constructs the Pad\'e $P_{(2,1)}$  luminosity distance and reconstructs the expansion history. Section~\ref{sec:data} describes the datasets and the statistical methodology. Finally, the results and interpretations are discussed in Section~\ref{sec:results}.

\section{Theoretical Framework of VCDM Model}\label{sec:vcdm}
The theoretical foundation of the VCDM model can be understood most clearly within the ADM (Arnowitt - Deser - Misner) $3+1$ decomposition of spacetime \cite{Arnowitt:1962hi}, which rewrites the metric as
\begin{equation}\label{Eq:1}
    ds^{2} = -N^{2} dt^{2} + \gamma_{ij}\,(dx^{i} + N^{i} dt)(dx^{j} + N^{j} dt),
\end{equation}
where $N$ is the lapse function, $N^{i}$ is the shift vector, and $\gamma_{ij}$ is the three-dimensional spatial metric on constant time hypersurfaces. Under the assumption of a spatially flat FLRW background, which describes a homogeneous and isotropic Universe, we consider the metric components as \(N=N(t)\), \(N^{i}=0\), and \(\gamma_{ij}=a^{2}(t)\delta_{ij}\). In this setup, the gravitational action can be expressed as 
\begin{equation}
    S = \int dt\, d^{3}x\, N \sqrt{\gamma}\, f(K_{ij}, R, N).
\end{equation} 
where $K_{ij}$, $R$ denotes the extrinsic curvature of the spatial hypersurface and the Ricci scalar associated with $\gamma_{ij}$, respectively \cite{DeFelice:2020eju,DeFelice:2020prd,DeFelice:2022uxv}. In Type-II MMG, the resulting auxiliary scalar mode generates a mild time dependence in the effective cosmological constant. Consequently, VCDM cosmology appears as the natural homogeneous and isotropic background realization of this ADM-based modification of gravity. Thus, the cosmic expansion rate in the VCDM framework is described by a modified Friedmann equation of the form \cite{DeFelice:2020cpt,Arnowitt:1962hi,Ganz:2022zgs}
\begin{equation} \label{eq:motH}
    \frac{H^{2}}{H_{0}^{2}} =
    \beta_{H}^{2} 
    + \Omega_{m,0}\!\left((1+z)^{3} - 1\right)
    + \Omega_{r,0}\!\left((1+z)^{4} - 1\right)
    + (1-\beta_{H}^{2})\,
    \frac{1 + \tanh\!\left(\dfrac{A_{2} - z}{A_{3}}\right)}
         {1 + \tanh\!\left(\dfrac{A_{2}}{A_{3}}\right)}.
\end{equation}
Here, $H_{0}$ is the present Hubble constant, $\Omega_{m,0}$ and $\Omega_{r,0}$ are the current matter and radiation density parameters, and the remaining terms encode the evolving vacuum energy component. The parameter $\beta_{H}$ measures the fractional deviation of the early time vacuum energy from the standard $\Lambda$CDM value.  It should be noted that when $\beta_{H} = 1$, the standard $\Lambda$CDM limit is recovered. 
Further, the constants $A_{2}$ and $A_{3}$ characterise the transition behaviour of the vacuum component \cite{DeFelice:2020cpt,Akarsu:2024qsi}.
\section{Cosmography in VCDM Model}
\label{sec:csmgrphy}
In the standard approach, the cosmographic parameters are treated as free quantities \cite{Visser:2004bf,Capozziello:2011hj}. In the present VCDM framework, we do not treat the present-day cosmographic parameters \(q_0\) and  \(j_0\) as independent free parameters. Instead, they are explicitly derived from the underlying VCDM model parameters \((\Omega_{m,0}, \beta_H, A_2)\). To establish the mathematical foundation, we present the Taylor series expansion of the scale factor around the present cosmic time \(t_0\)
\begin{equation}
a(t) = a(t_{0})+ \dot{a}(t_{0})(t - t_{0})+ \frac{1}{2!}\,\ddot{a}(t_{0})(t - t_{0})^{2}+ \frac{1}{3!}\,a^{(3)}(t_{0})(t - t_{0})^{3}+ \frac{1}{4!}\,a^{(4)}(t_{0})(t - t_{0})^{4}+ \cdots ,
\end{equation}
where $a^{(n)}(t)$ denotes the $  n  $-th time derivative of the scale factor $a(t)$ \cite{Visser:2004bf,Cattoen:2007sk}. Truncating the expansion at fourth order (i.e., neglecting terms of order \(\mathcal{O}[(t-t_0)^4]\) and higher) naturally introduces the cosmographic parameters in the time domain as follows
\begin{equation}
H(t) = \frac{\dot{a}(t)}{a(t)},\quad
q(t) = -\,\frac{a(t)\,\ddot{a}(t)}{\dot{a}(t)^{2}},\quad
j(t) = \frac{a(t)^{2}\,\dddot{a}(t)}{\dot{a}(t)^{3}}.\quad
\end{equation}
Here, \(H(t)\) is the Hubble parameter, \(q(t)\) is the deceleration parameter and, \(j(t)\) is the jerk parameter. For comparison with observational data, it is convenient to recast these quantities as functions of redshift \(z\). Using the relations,
$\displaystyle 1 + z = \frac{a_0}{a(t)} \; \text{ and} \; \frac{d}{dt} = -H(1+z)\frac{d}{dz}$,
the  explicit expressions of the cosmographic parameters \(H(z)\), \(q(z)\), and \(j(z)\) in terms of redshift derivatives of \(H(z)\) are given by
\begin{align}
q(z) =& -1 + (1 + z)\frac{{H}'(z)}{H(z)}, \label{eq:motcgpy} \\
j(z) =& 1 - 2(1 + z)\frac{{H}'(z)}{H(z)} + (1 + z)^2\left(\frac{{H}''(z)}{H(z)} + \left(\frac{{H}'(z)}{H(z)}\right)^2\right), \label{eq:motcgpyI} 
\end{align}
where the prime denotes differentiation with respect to the redshift \(z\) \cite{Gruber:2013wua,Visser:2004bf}.  In the VCDM model, the Hubble function \(H(z)\) is fully determined by the model parameters \((\Omega_{m,0}, \beta_H, A_2)\) (see Eq.~\eqref{eq:motH}) \cite{DeFelice:2020cpt,Akarsu:2024eoo}. Consequently, the remaining free parameters to be constrained from observational data are $(\Omega_{m,0}, \beta_H, A_2)$. Now substituting the successive redshift derivatives \({H}'(z)\), \({H}''(z)\), \({H}'''(z)\) in Eqs.~(\ref{eq:motcgpy})--(\ref{eq:motcgpyI}) and evaluating at the present epoch (\(z = 0\)) yields
\begin{align}
q_{0}&= -1 + \frac{1}{2}\left[3\Omega_{m,0}-\left(1-\beta_{H}^{2}\right)\frac{S}{A_{3}D}\right], \label{eq:q0_def} \\[1ex]
j_{0}&= 1+ (1-\beta_{H}^2)\frac{S\left(A_3+T\right)}{D A_3^2}, \label{eq:j0_def} 
\end{align}
where the quantities $T$, $S$, and $D$ are defined as
\begin{equation*}
T = \tanh\left(\frac{A_2}{A_3}\right), \quad
S = \operatorname{sech}^2\left(\frac{A_2}{A_3}\right), \quad
D = 1 + T.
\end{equation*}
Further, the functions $F$ and $G$ are given by
\begin{displaymath}
F(A_2,A_3,\Omega_{m,0}) =\frac{S}{D}\left[\frac{3}{2A_3}+ \frac{3T}{A_3^2}+ \frac{2T^2 - 1}{A_3^3}\right]- \frac{3\Omega_{m,0} S}{A_3 D},\, G(A_2,A_3) =\frac{S^2}{4 A_3^2 D^2}.    
\end{displaymath}
The expressions in  Eqs.~(\ref{eq:q0_def})--(\ref{eq:j0_def}) show that deviations from these \(\Lambda\)CDM values are directly controlled by the factors \((1 - \beta_H^2)\) , \(A_2\), and \(A_3\), respectively.
\section{Pad\'e  Cosmographic Reconstruction of the Luminosity Distance}
\label{sec:pade} The Pad\'e approximants offer superior global approximation properties compared to the truncated Taylor series and remain well-behaved over a wider redshift range as discussed in Section~\ref{intro}. The standard Pad\'e (2,1) form for the luminosity distance, matched to the cosmographic expansion up to the jerk order is given by \cite{Hu:2022udt}
\begin{equation}
d_L(z)=\displaystyle\frac{cz\left[\left(-2j_0 + q_0(3q_0 + 8) - 5\right)z + 6(q_0 - 1)\right]}{H_0\left[2q_0\left(3q_0 z + z + 3\right)-2\left(j_0 z + z + 3\right)\right]}.
\label{eq:dl_general}
\end{equation}
A detailed description of these rational approximants can be found in \cite{Capozziello:2020ctn}.
Substituting the VCDM-derived expressions Eqs.~(\ref{eq:q0_def}) and (\ref{eq:j0_def}) into the above Pad\'e (2,1) approximant yields a compact analytic expression for the luminosity distance fully determined by the fundamental VCDM model parameters set $\{\beta_H,\Omega_{m,0},A_2,A_3\}$, which is expressed as follows
\begin{equation}
d_L(z)=\frac{c}{H_0}\frac{z\left[\left(-2J + Q(3Q+8) - 5\right)z + 6(Q - 1)\right]}{2Q(3Qz + z + 3) - 2(Jz + z + 3)},
\label{eq:dl_compact}
\end{equation}
where 
\begin{align*}
Q \equiv -1 + \frac{1}{2}\left[3\Omega_{m,0}- \frac{(1-\beta_H^2)}{A_3}\frac{S}{D}\right] \;\text{and}\;J \equiv 1 +\frac{(1-\beta_H^2)\, S \, (A_3 + T)}{D A_3^2}.
\label{eq:J_def}
\end{align*}
The Hubble parameter \(H(z)\) can then be reconstructed from the luminosity distance using the standard relation
\begin{equation}\label{Eq:14}
H(z) = c \left[ \frac{d}{dz} \left( \frac{d_L(z)}{1+z} \right) \right]^{-1}.
\end{equation}
Implementing the general Pad\'e (2,1) expression \eqref{eq:dl_general} into Eq.~\eqref{Eq:14} gives
\begin{equation}\label{eq:H_gen}
H(z) = 2H_0 (1+z)^2 \displaystyle\frac{[N_g(z)]^2}{D_g(z)},
\end{equation}
where
\begin{align*}
N_g(z) &= j_0 z - q_0 (3 q_0 z + z + 3) + z + 3, \\[1ex]
D_g(z) &= z^2 \left(2 j_0^2 + j_0 (7 - 9 q_0^2 - 10 q_0) + q_0 (9 q_0^3 + 18 q_0^2 + 17 q_0 - 40) + 14 \right) \nonumber \\
&\quad + 6(q_0 - 1) z \left[-2 j_0 + q_0 (3 q_0 + 8) - 5\right] + 18(q_0 - 1)^2.
\end{align*}
Finally, by using Eqs.~\eqref{eq:q0_def}-\eqref{eq:j0_def}, $H(z)$ in terms of the model parameters can be obtained as follows
\begin{equation}\label{Eq:16}
H(z) = 2H_0 (1+z)^2 \frac{[N_g(z;\Omega_{m,0},\beta_H,A_2,A_3)]^2}{D_g(z;\Omega_{m,0},\beta_H,A_2,A_3)}.
\end{equation}
The aforementioned Pad\'e  $P_{(2,1)}$ reconstruction now provides a robust analytic form of \(H(z)\) within the VCDM framework, which is suitable for direct comparison with late-time observational data. 

\section{Datasets and Statistical Methodology}
\label{sec:data}
To test the viability of the VCDM scenario within the Pad\'e $P_{(2,1)}$ cosmographic framework,
we perform a joint analysis using several late-time cosmological probes that are
sensitive to both the expansion history and its redshift evolution.
The adopted datasets are chosen to ensure complementarity in the measured observables
and to minimize potential overlaps in systematic uncertainties.
In particular, we combine cosmic chronometer measurements, DESI BAO data, and Type-Ia supernova observations, which together provide a coherent and robust
description of the late-time Universe. Cosmic chronometers directly constrain the Hubble expansion rate, DESI BAO measurements anchor absolute distance scales through characteristic clustering features, and supernova data trace the integrated luminosity distance over a wide
redshift range. The simultaneous use of these probes allows us to break parameter degeneracies and to tightly constrain the cosmographic parameters predicted by the VCDM model.

The parameter estimation is carried out within a Bayesian framework using the Markov Chain Monte Carlo (MCMC) technique. In this approach, samples are drawn from the posterior distribution by constructing a Markov chain that asymptotically converges to the target posterior. For this purpose, we employ the affine-invariant ensemble sampler implemented in the \texttt{emcee} Python package~\cite{Foreman-Mackey:2012any}, which is particularly efficient in exploring high-dimensional parameter spaces with non-trivial degeneracies.
\subsection{Cosmic Chronometers}

The cosmic chronometer (CC) technique allows one to estimate the Hubble parameter $H(z)$ directly from the differential ageing of passively evolving galaxies \cite{Jimenez:2001gg}. As a result, this method relies on the observable quantity $dz/dt$ and provides a model-independent test of the late-time cosmic expansion without assuming any external distance calibration. In this study, we use a sample of 34 CC measurements spanning the redshift range $0<z<2$, compiled from several observational sources \cite{Moresco:2016mzx, Moresco:2020fbm}. The dataset is divided into two statistically distinct parts. First, we consider 15 measurements from the homogeneous compilation, together with its publicly available covariance matrix\footnote{\url{https://gitlab.com/mmoresco/CCcovariance}} that accounts for correlated systematic uncertainties. Second, we include 19 independent CC measurements obtained from various observational studies \cite{Moresco:2020fbm}, and treat them as mutually uncorrelated with their reported errors. Therefore, the total likelihood is constructed as the sum of correlated and uncorrelated contributions,
\begin{equation}
\chi^2_{\mathrm{CC}}=\chi^2_{\mathrm{corr}}+\chi^2_{\mathrm{uncorr}}.
\end{equation}
The correlated component is evaluated using the full covariance matrix, whereas the remaining measurements are incorporated through the standard $\chi^2$ expression with a diagonal error matrix \cite{Kavya:2025vsj}. CC measurements provide strong constraints over the redshift range $z\lesssim1.5$, and can therefore be used to test the background evolution and parameter space of the VCDM model when combined with DESI DR2 BAO and Type Ia supernova observations.

\subsection{DESI BAO DR2}

The Dark Energy Spectroscopic Instrument (DESI) Data Release 2 (DR2) provides the latest and most precise BAO measurements from galaxies, quasars, and the Lyman-$  \alpha  $ forest \cite{DESI:2025zgx}. The dataset covers a wide redshift range $0.295 \lesssim z \lesssim 2.33$ and includes the transverse comoving distance $  D_M(z)  $, Hubble distance $  D_H(z)  $, and their ratios to the sound horizon $  r_d  $, together with the full covariance matrix. We adopt the DESI DR2 BAO measurements and covariance in all four dataset combinations (CC + DESI, supplemented by Pantheon+, Union3, or DES-SN5YR). The BAO data provide strong constraints on the expansion history at intermediate and high redshifts, as demonstrated in the DESI cosmological analyses \cite{DESI:2024mwx}, making them particularly effective for testing the smoothness of the late-time transition and the jerk parameter predicted by the VCDM model.
\subsection{Supernovae}
Type Ia supernovae remain the most direct and powerful probe of the late-time cosmic expansion history through their standardized luminosity distances. In this work, we employ three major supernova compilations, such as Pantheon+, Union3, and DES-SN5YR, and each is supplemented with the cosmic chronometers and DESI BAO DR2 data. The datasets and the respective covariance matrices can be found in the repository\footnote{\url{https://github.com/CobayaSampler/sn_data}}.

\paragraph{\textbf{Union3:}}
The Union3 compilation \cite{Rubin:2023jdq} combines 2087 spectroscopically confirmed Type Ia supernovae from multiple surveys, extending up to $  z \sim 1.5  $. It offers improved control over systematics and a homogeneous sample suitable for cosmology. We use the released distance moduli together with the full covariance matrix in the combination CC + DESI + Union3 \cite{Rubin:2023jdq, Mishra:2025vpy, SupernovaCosmologyProject:2015zlj}.
\paragraph{\textbf{Pantheon+:}}
The Pantheon+ compilation \cite{Brout:2022vxf,Scolnic:2021amr}
is the largest publicly available catalog of Type Ia supernovae, consisting of 1701 spectroscopically confirmed events with well-calibrated distance moduli. It spans a wide redshift range up to $  z \sim 2.3  $, providing strong leverage on both low- and high-redshift expansion history. We adopt the distance moduli and the full covariance matrix, which already marginalise all major systematic uncertainties, and include it in our dataset combinations (CC + DESI + Pantheon+).
\paragraph{\textbf{DES-SN5YR}}
The DES-SN5YR sample \cite{DES:2024ffp} provides one of the largest homogeneous sets of photometrically classified Type Ia supernovae, with a total cosmology-grade sample extending to $  z \sim 1.3  $. Its high-redshift leverage is particularly useful for probing the late-time expansion. We adopt the distance moduli and the full covariance matrix in the combination CC + DESI + DES-SN5YR \cite{DES:2025tir}.
\section{Interpretations and Final Remarks}
\label{sec:results}
\begin{table}
\centering
\small 

\caption{
Constraints up to $1\sigma$ on the model parameters obtained from different dataset combinations.
}
\begin{tabular}{|c|c|c|c|c|c|}
\hline 
Dataset & $H_0$ & $\Omega_{m,0}$ & $\beta_H$ & $A_2$ & $r_d$ \\[1ex]
\hline
CC+DESI & $68.3\pm2.1$ & $0.3727\pm0.0087$ & $1.00\pm0.27$ & $0.39_{-0.32}^{+0.15}$ & $145.0\pm 4.3$ \\[1ex] 
CC+DESI+Union3 & $69.6\pm1.7$ & $0.3726 \pm 0.0082$ & $0.999\pm0.20$ & $0.33_{-0.21}^{+0.15}$ & $142.3\pm3.4$ \\[1ex] 
CC+DESI+PantheonPlus & $72.58\pm0.20$ & $0.3702 \pm 0.0080$ & $0.997\pm0.18$ & $0.31_{-0.15}^{+0.13}$ & $136.66\pm0.96$ \\[1ex]
CC+DESI+DESY5 & $69.17\pm0.20$ & $0.3720 \pm 0.0075$ & $1.00\pm0.20$ & $0.32_{-0.21}^{+0.15}$ & $143.14\pm0.79$ \\
\hline
\end{tabular}

\label{tab:tab1}
\end{table}
\begin{figure}
    \centering
    \includegraphics[width=0.85\linewidth]{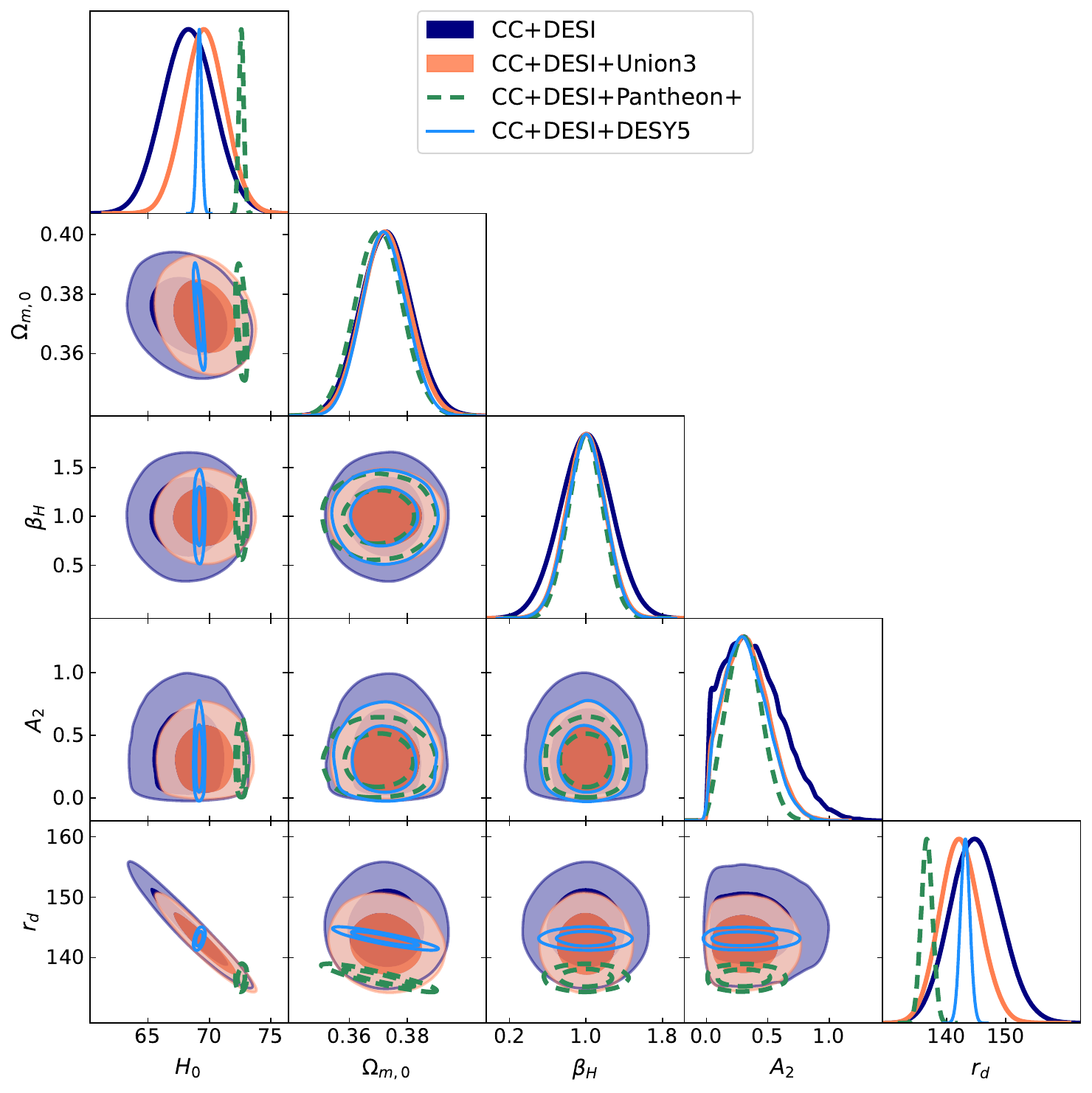}
    \caption{\justifying
Joint posterior distributions of the parameter space \{$H_0$, $\Omega_{m,0}$, $\beta_H$, $A_2$, $r_d$\} for different dataset combinations. The contours correspond to the $1\sigma$ and $2\sigma$ confidence regions. Different colors represent the combinations CC+DESI, CC+DESI+Union3, CC+DESI+Pantheon+, and CC+DESI+DESY5. The results show consistent constraints across datasets, with mild shifts primarily driven by the inclusion of supernova data.
}
    \label{fig:maincontour}
\end{figure}
\begin{figure}
    \centering
    \includegraphics[width=0.75
    \linewidth]{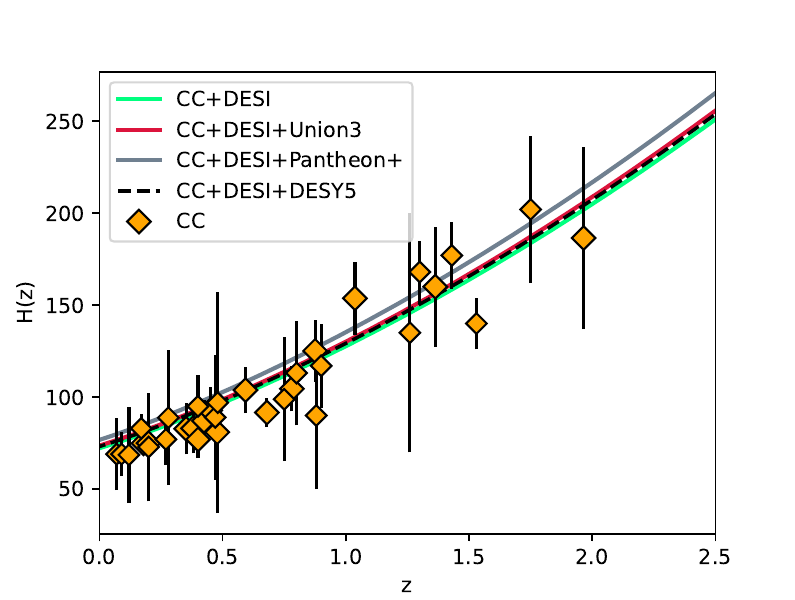}
    \caption{\justifying Evolution of the Hubble parameter $H(z)$ for different dataset combinations. The cosmic chronometer data points are shown for comparison. All models exhibit consistent behavior over the redshift range, with only minor variations in normalization.}
    \label{fig:h}
\end{figure}
\begin{figure}
    \centering
    \includegraphics[width=0.75
    \linewidth]{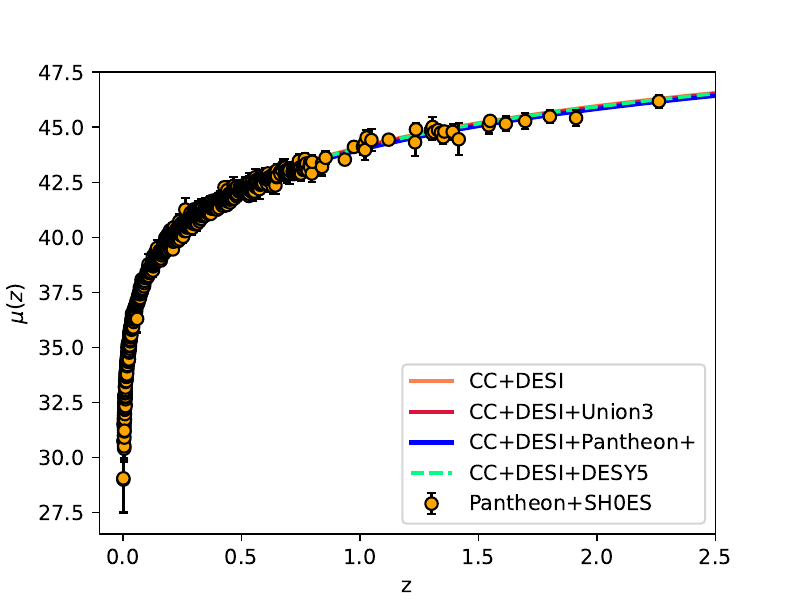}
    \caption{
Distance modulus $\mu(z)$ as a function of redshift for different dataset combinations. The theoretical predictions are compared with the Pantheon+SH0ES supernova data. The model provides an excellent fit to the observational data across the full redshift range, with negligible differences among the dataset combinations.
}
    \label{fig:mu}
\end{figure}
\begin{figure}
    \centering
    \includegraphics[width=0.65\linewidth]{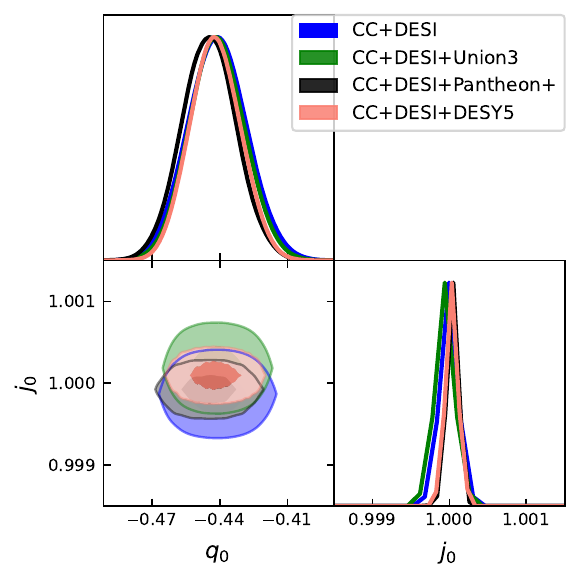}
    \caption{\justifying 
Constraints in the $q_0$-$j_0$ plane for different dataset combinations. The contours represent the $1\sigma$ and $2\sigma$ confidence regions. All datasets yield values of the jerk parameter tightly clustered around $j_0 \approx 1$, indicating no significant deviation from the $\Lambda$CDM prediction at the level of third-order cosmography. The spread is primarily along the $q_0$ direction, reflecting the sensitivity of the data to the present cosmic acceleration.
}
    \label{fig:qj}
\end{figure}
\begin{table}
\centering
\caption{Constraints on the deceleration parameter $q_0$ and jerk parameter $j_0$ for different dataset combinations (1$\sigma$ confidence level).}
\begin{tabular}{|c|c|c|}
\hline
\textbf{Dataset} & \boldmath{$q_0$} & \boldmath{$j_0$} \\
\hline
CC+DESI 
& $-0.441 \pm 0.013$ 
& $0.999986^{+0.000015}_{-0.000013}$ \\

CC+DESI+Union3 
& $-0.441 \pm 0.012$ 
& $1.000004^{+0.000003}_{-0.000005}$ \\

CC+DESI+Pantheon+ 
& $-0.445 \pm 0.012$ 
& $0.999993^{+0.000008}_{-0.000006}$ \\

CC+DESI+DESY5 
& $-0.442 \pm 0.011$ 
& $0.999988^{+0.000013}_{-0.000011}$ \\
\hline
\end{tabular}
\label{tab:q0j0_constraints}
\end{table}

In the spectrum of alternatives to the GR framework, it is very difficult to find a theory that explains all aspects of the Universe. The latest observations reveal loopholes in the standard framework. This has led researchers to move beyond the GR+$\Lambda$CDM background \cite{Verde2019Tension} and to probe the dynamics of the Universe through alternative theories. In pursuit of this quest, we explore a theory that has shown promising directions in explaining the evolution of the Universe \cite{DeFelice:2021xps,DeFelice:2022riv,DeFelice:2020onz,DeFelice:2022uxv,Jalali:2023wqh,Ganz:2024ihb} .
Particularly, we studied the late-time expansion history of the Universe within the VCDM model, a realization of Type-II MMG in which the vacuum sector evolves dynamically without introducing additional propagating degrees of freedom beyond those of GR. By combining the exact background solution with a Pad\'e $P_{(2,1)}$ cosmographic reconstruction, we obtained stable predictions for both the Hubble parameter and the luminosity distance across the full redshift range currently accessible to observations, including the Pantheon$+$ supernova sample up to $z \approx 2.3$.

\paragraph{\textbf{Parameter Constraints:}}

By performing joint MCMC analyses with four dataset combinations, CC+DESI, CC+DESI+Union3, CC+DESI+Pantheon$+$, and CC+DESI+DESY5, we found mutually consistent constraints on the VCDM parameter space. The resulting posterior distributions, summarised in ~\autoref{tab:tab1} and visualised in \autoref{fig:maincontour}, confirm the robustness of the parameter estimates across all combinations. The vacuum parameter $\beta_H$ remains close to unity throughout, indicating that the $\Lambda$CDM limit is comfortably contained within the preferred region of parameter space. For the transition parameter $A_2$, the posterior distributions show a mild tendency toward positive values, typically around $0.31$--$0.39$. However, $A_2 = 0$ remains allowed at roughly the $\lesssim 2\sigma$ level for all dataset combinations. Therefore, current background data do not provide statistically significant evidence for vacuum evolution in the VCDM framework. The inferred sound horizon $r_d$ displays the expected dataset dependence. Inclusion of Pantheon$+$ \cite{Scolnic:2021amr} shifts the fit toward a higher Hubble constant, $H_0 = 72.58 \pm 0.20$ km s$^{-1}$ Mpc$^{-1}$, together with a lower sound horizon, $r_d = 136.66 \pm 0.96$ Mpc. The remaining combinations prefer $H_0 \approx 68$--$70$ km s$^{-1}$ Mpc$^{-1}$ and $r_d \approx 142$--$145$ Mpc. This behavior is naturally interpreted as part of the well-known $H_0$--$r_d$ tension between early- and late-time probes \cite{Riess2022SH0ES,Verde2019Tension} rather than as evidence of internal inconsistency in the VCDM scenario.  The present matter density ($\Omega_{m,0}$) remains tightly constrained in all cases.

\paragraph{\textbf{Hubble Expansion and Distance Modulus Reconstruction:}} The reconstructed Hubble function $H(z)$ provides an adequate fit to all 34 cosmic chronometer measurements \cite{Moresco:2020fbm} over $0 < z < 2$ as shown in ~\autoref{fig:h}. Up to intermediate redshift, $z \lesssim 0.8$, all four dataset combinations yield nearly identical expansion histories. Differences become visible only at higher redshift and are driven primarily by the larger $H_0$ value preferred when Pantheon$+$ data are included. A similar picture emerges for the distance modulus. In ~\autoref{fig:mu}, the reconstructed curves remain in excellent agreement with the Pantheon$+$SH0ES sample across the full redshift range, with all four fits overlapping at the sub-percent level. This supports the use of the Pad\'e $P_{(2,1)}$ approximation as a robust alternative to standard Taylor cosmography \cite{Gruber:2013wua,Capozziello:2017ddd}, particularly at $z \gtrsim 1$ where Taylor expansions are known to suffer from convergence problems \cite{Cattoen:2007sk} and systematic bias.

\paragraph{\textbf{Cosmographic Parameters and the Jerk Diagnostic:}} The inferred cosmographic parameters further clarify the late-time behaviour of the model. For all dataset combinations, the present deceleration parameter is close to $q_0 \approx -0.44$, confirming that the Universe is undergoing accelerated expansion today. The numerical constraints on $q_0$ and $j_0$ are listed in ~\autoref{tab:q0j0_constraints} (up to 1$\sigma$) and can be visualized in~\autoref{fig:qj} (up to 2$\sigma$). More importantly, the jerk parameter is consistently found to be $j_0 \approx 1$, in agreement with the $\Lambda$CDM prediction at the $10^{-3}$ level \cite{Visser2004Cosmography}. This is nontrivial in light of recent model-independent cosmographic reconstructions based on DESI DR2 BAO \cite{Rodrigues:2025tfg}, some of which reported significant departures from $j_0 = 1$. Within VCDM, this quantity is not an independent fitting parameter. The deviation $\Delta j_0 = j_0 - 1$ is determined by the underlying theory through the parameters $A_2$, $\Omega_{m,0}$, and the fixed scale $A_3 = 10^{-3}$. In the region preferred by our data, where $\beta_H \approx 1$ and $A_2 \lesssim 0.4$, the factor $(1-\beta_H^2)$ strongly suppresses any deviation from unity, resulting in $|\Delta j_0| \lesssim 10^{-4}$. The jerk parameter provides a direct internal consistency test of the model. If future cosmographic analyses establish a statistically significant deviation from $j_0 = 1$, then the viable VCDM parameter space would be significantly reduced. Such a result would require either a substantial departure of $\beta_H$ from unity or considerably larger values of $A_2$, both of which are disfavored by the present joint constraints. In that sense, VCDM differs qualitatively from phenomenological jerk parameterizations \cite{Rapetti:2006fv} in which $j_0$ is introduced only as a free kinematic quantity.

\paragraph{\textbf{Physical Implications:}} The present results indicate that, at the background level, VCDM remains observationally close to $\Lambda$CDM, similar to other modified gravity scenarios constructed to mimic the concordance expansion history while modifying the underlying dynamics \cite{Kolhatkar:2026ixl, Mishra:2023onl}. This is not a weakness of the model; rather, it reflects the present precision frontier of background observables. A viable modified gravity theory is not required to generate large deviations from $\Lambda$CDM at all scales or epochs. Instead, it must remain consistent with existing data while allowing testable departures in regimes where precision is sufficient. From that perspective, the near-unity values of $\beta_H$ and $j_0$ are informative. They imply that any late-time modification of the vacuum sector, if present, must be modest within the range probed by current background observations. The strongest leverage on the model is therefore unlikely to come from background expansion data alone. 
In the direct confrontation of the VCDM model with observational data, a transition feature in the Hubble expansion rate, as reported in earlier works, is clearly observed \cite{DeFelice:2020cpt, Mishra:2026vdo}. However, when the same framework is constrained through the  Pad\'e $P_{(2,1)}$ cosmographic reconstruction, this transition is no longer present. This indicates that the transition is not a robust feature required by the data, but rather depends on the specific parametrization of the model. In particular, the constraints obtained in this work drive the model parameters toward a regime where deviations from $\Lambda$CDM are strongly suppressed, leading to a smooth expansion history without any distinct transition. This suggests that current observational data do not provide compelling evidence for the vacuum metamorphosis transition, and instead favor a cosmological evolution closely resembling the standard $\Lambda$CDM scenario.

\paragraph{\textbf{Future Outlook:}} A more decisive test of VCDM will require substantially tighter bounds on the parameter $A_2$ or the inclusion of perturbation-level observables sensitive to structure formation and gravitational response. These include CMB anisotropies, weak lensing, redshift-space distortions, and growth-rate measurements, where cosmological perturbations can reveal signatures hidden at the background level \cite{Kolhatkar:2025ubm}. Such probes can distinguish modified Friedmann dynamics even when the background expansion history closely mimics $\Lambda$CDM. The embedding of the $\Lambda_{\rm s}$CDM sign-switching scenario within the VCDM framework, together with the reported compatibility of the model with neutrino-mass bounds, further motivates a broader phenomenological study \cite{Arora:2025msq,Ladeira:2026jne}. Beyond the observational probes considered in the present work, the Pad\'e cosmographic framework can naturally be extended to future high-redshift measurements. In particular, gravitational-wave standard sirens \cite{LIGOScientific:2017adf} detected by third-generation observatories, such as the Einstein Telescope \cite{Punturo:2010zz} and Cosmic Explorer \cite{Reitze:2019iox}, as well as the space-based LISA mission \cite{LISA:2017pwj}, will provide independent luminosity-distance measurements over a much wider redshift range without relying on the traditional cosmic distance ladder. Since the Pad\'e reconstruction remains well behaved at high redshift, unlike conventional Taylor cosmography, it offers a promising framework for incorporating these standard sirens into model-independent tests of the cosmic expansion history. Such complementary observations have the potential to significantly enhance the capability to distinguish modified gravity scenarios from dynamical dark-energy models and other forms of new physics by simultaneously probing both the background expansion and, when combined with structure-growth measurements, the gravitational sector itself. Moreover, future joint analyses including gravitational-wave observations, CMB, weak lensing, redshift-space distortions, and forthcoming DESI, Euclid \cite{Linke:2024zdt}, and LSST \cite{LSSTDarkEnergyScience:2018yem} data may considerably improve the determination of the cosmographic parameters and provide a more decisive assessment of whether VCDM can alleviate the existing $H_0$ and $S_8$ tensions or whether its background evolution remains observationally indistinguishable from the concordance $\Lambda$CDM model. Taken together, these future observational probes and perturbation-level analyses will provide a significantly sharper assessment of whether VCDM is merely a $\Lambda$CDM-like limit of modified gravity or a genuinely distinguishable description of cosmic acceleration.

\section*{Data availability} No new data was generated or analyzed to support this research.

\section*{Acknowledgment} SKB acknowledges the Council of Scientific and Industrial Research (CSIR), Govt. of India for awarding Junior Research fellowship (E-Certificate No.: 22D/23J06347). PKS acknowledges IUCAA, Pune, India for providing support through the visiting Associateship program. 
We are very grateful to the honorable referees and the editor for the illuminating suggestions that have significantly improved our work in terms of research quality and presentation.

\bibliography{main}

\end{document}